# Chloride-induced alterations of the passive film on 316L stainless steel and blocking effect of pre-passivation

Zuocheng Wang, Antoine Seyeux, Sandrine Zanna, Vincent Maurice[1], Philippe Marcus[2]
*PSL Research University, CNRS - Chimie ParisTech, Institut de Recherche de Chimie Paris, Research Group Physical Chemistry of Surfaces, Paris, France*

## Abstract

Electrochemical polarization measurements were combined with surface analysis by Time of Flight Secondary Ion Mass Spectroscopy (ToF-SIMS), X-Ray Photoelectron Spectroscopy (XPS) and Atomic Force Microscopy (AFM) to study the alterations of the passive film on 316L austenitic stainless steel induced by the presence of chlorides in sulfuric acid electrolyte. The work was performed at a stage of initiation of localized corrosion preceding metastable pitting at the micrometer scale as verified by current transient analysis and AFM. The results show that $Cl^-$ ions enter the bilayer structure of the surface oxide already formed in the native oxide-covered initial surface state at concentrations below the detection limit of XPS (< 0.5 at%), mostly in the hydroxide outer layer where Fe(III) and Mo(IV,VI) species are concentrated but barely in the oxide inner layer enriched in Cr(III). Their main effect is to produce a less resistive passive state by poisoning dehydroxylation and further Cr(III) and Mo(IV,VI) enrichments obtained in the absence of chlorides. This detrimental effect can be suppressed by pre-passivation in a Cl-free electrolyte, which blocks the entry of chlorides in the passive film, including in the outer exchange layer, and enables the beneficial aging-induced variations of the composition to take place despite the presence of chlorides in the environment.

## Keywords

[1] Corresponding author : vincent.maurice@chimieparistech.psl.eu

[2] Corresponding author : philippe.marcus@chimieparistech.psl.eu

## Introduction

Passive films consisting of ultrathin oxide/hydroxide layers formed at the surface make stainless steels (SS) highly resistant against corrosion, even in harsh environmental conditions. On ferritic Fe-Cr(-Mo) [1-19] and austenitic Fe-Cr-Ni(-Mo) [20-40] SS substrates, the surface oxide/hydroxide layer is only a few nanometers thick but markedly enriched in Cr(III), which is the key for efficient passivity because of the higher stability of Cr(III) compared to Fe(II,III) oxide species. For austenitic SS, only very little Ni(II) is present when detected in the passive film and the metallic alloy region underneath the oxide is enriched with Ni(0) [20-22,27-30].

With the addition of Mo in the alloy, e.g. in austenitic AISI 316L, an increased corrosion resistance is obtained in chloride-containing environments, where passive film breakdown can be followed by the initiation of localized corrosion by pitting. The passive films are similarly ultrathin but the composition include Mo(IV) or Mo(VI) oxide species at a few at% level [10,21,22,30,32,35,37]. It has been proposed, and it is still debated, that the beneficial role of molybdenum would be to mitigate passive film breakdown [10,19,28,30,36,39,40-45] or to promote passive film repair [18,30,32,33,41,43-45].

As suggested by recent nanometer scale studies [27,48], the Cr(III) enrichment may not be homogeneous in the passive film, and the Cr enrichment heterogeneities may cause the local failure of the passivity and the initiation of localized corrosion followed by pit growth where the passive film fails to self-repair [49]. The better understanding of the mechanisms governing the Cr (and Mo) enrichment requires to thoroughly investigate the initial stages of oxidation leading to pre-passivation of the SS surface [50,51,52] as well as the alterations brought by electrochemical passivation of the native oxide-covered SS surface [46,47].

On AISI 316L, recent surface analytical studies performed by Time-of-Flight Secondary Ion Mass Spectroscopy (ToF-SIMS) and X-ray Photoelectron Spectroscopy (XPS) [27,46,47] confirmed the bilayer structure of the passive film previously reported [21,22,27,30,32]. The native oxide film formed in air was found to already develop this bilayer structure with both the hydroxide outer and oxide inner layers enriched in Cr(III) and with Fe(III) more concentrated in the outer layer together with Mo(IV,VI). Electrochemical passivation in chloride-free sulfuric acid solution did not alter the bilayer structure and thickness of the surface oxide but promoted its Cr and Mo enrichments, owing to the lower stability and preferential dissolution of Fe(III) as previously proposed [4,16,53], and thereby increased the corrosion resistance of the passive state.

In the present work, we address the alterations of the surface oxide films brought by electrochemical passivation in Cl-containing sulfuric acid solutions. Potentiodynamic and potentiostatic polarization measurements were used to define the electrochemical conditions best suited to alter the passive state without initiating localized corrosion (i.e metastable pitting) at the micrometer scale. ToF-SIMS and XPS were applied to characterize the bilayer structure, thickness and composition of the passive film and the entry of chlorides. Surface morphology was studied by Atomic Force Microscopy (AFM). The results provide new insight on the effect of pre-passivation on the entry of chlorides in the passive film and increased resistance to metastable pitting at the nanometer scale.

## Experimental

The same polycrystalline AISI 316L austenitic SS samples as in previous work were used [46,47]. The bulk composition in the main alloying elements was Fe–19Cr–13Ni–2.7Mo wt% (Fe–20Cr–12Ni–1.6Mo at%). The surface was prepared by mechanical polishing first with emery paper of successive 1200 and 2400 grades and then with diamond suspensions of successive 6, 3, 1 and 0.25 μm grades. Cleaning and rinsing were performed after each polishing step in successive ultrasonicated baths of acetone, ethanol and Millipore® water (resistivity > 18 MΩ cm). Filtered compressed air was used for drying.

A 3-electrode electrochemical cell controlled by a Gamry 600 potentiostat was used for the electrochemical measurements. It included a Pt grid as counter electrode and a saturated calomel electrode as reference electrode. The area of the working electrode was 0.5 $cm^2$ delimited by a Viton O-ring. The electrolyte was a 0.05 M $H_2SO_4$ aqueous solution in which NaCl was added at varying concentration. It was prepared from ultrapure chemicals (VWR®) and Millipore® water and bubbled with argon for 30 minutes prior to and continuously during the measurements. All experiments were performed at room temperature.

Passivation in the Cl-containing electrolyte was performed starting from two different surface states previously analyzed in details [46]: the native oxide-covered surface state and the pre-passivated surface state. The native oxide-covered sample was obtained after a short time (5-10 min) exposure in ambient air after surface preparation. The pre-passivated sample was produced by potentiostatic anodic polarization in the passive domain in the Cl-free 0.05 M $H_2SO_4$ electrolyte: after resting for 0.5 hour at open circuit value ($U_{OC}$ = -0.23 V/SCE), the potential was stepped to $U_{Pass}$ = 0.3 V/SCE and maintained at this value for 1 hour. The $U_{Pass}$ value was selected from potentiodynamic polarization as described below. No cathodic pre-treatment was performed in order to avoid any reduction-induced alteration of the initial native oxide film prior to passivation. These exact same conditions as in our previous study were adopted so as to enable the comparative analysis of the effects brought by the presence of chlorides in the electrolyte.

In the Cl-containing electrolyte, two different passivated samples were prepared in order to study the alterations of the passive films brought by passivation in the presence of chloride and the modifications induced by pre-passivation of the surface. One sample was passivated directly from the initial native oxide-covered surface state using the same treatment conditions as for pre-passivation but in a 0.05 M $H_2SO_4$ + 0.05 M NaCl electrolyte. This concentration was selected from potentiostatic polarization curves as discussed below. The other sample was passivated in the same conditions but starting from the pre-passive surface state obtained in the Cl-free electrolyte.

Depth profile elemental analysis of the oxide-covered surfaces was performed using a ToF-SIMS 5 spectrometer (Ion ToF – Munster, Germany) operated at about $10^{-9}$ mbar. Topmost surface analysis in static SIMS conditions using a pulsed 25 keV $Bi^+$ primary ion source delivering 1.2 pA current over a 100 × 100 $\mu m^2$ area was interlaced with sputtering using a 1 keV $Cs^+$ ion beam giving a 32 nA target current over a 300 × 300 $\mu m^2$ area. Analysis was centered inside the eroded crater to avoid edge effects. The profiles were recorded with negative secondary ions that have higher yield for oxide matrices than for metallic matrices. The Ion-Spec software was used for data acquisition and processing.

Surface chemical analysis was performed by XPS with a Thermo Electron ESCALAB 250 spectrometer operating at about $10^{-9}$ mbar. The X-ray source was an $AlK_\alpha$ monochromatized radiation (hν = 1486.6 eV). Survey spectra were recorded with the pass energy of 100 eV at a step size of 1 eV. High resolution spectra of the Fe 2p, Cr 2p, Ni 2p, Mo 3d, O 1s, S 2p, Cl 2p and C 1s core level regions were recorded with a pass energy of 20 eV at a step size of 0.1 eV. The take-off angle of the analyzed photoelectrons was 90°. The binding energies (BE) were calibrated by setting the C 1s signal corresponding to olefinic bonds ($–CH_2–CH_2–$) at 285.0 eV. Data processing (curve fitting) was performed with CasaXPS, using a Shirley type background and Lorentzian/Gaussian (70%/30%) peak shapes. Asymmetry was taken into account for the metallic components ($Cr^0$, $Fe^0$, $Mo^0$, $Ni^0$) and a broader envelope was used to account for the multiplet splitting of the oxide components ($Cr^{III}$, $Fe^{III}$, $Mo^{IV}$, $Mo^{VI}$).

AFM analysis was performed with a Nano-Observer microscope (CSI instruments) operated in oscillating (Tapping®) mode. A "Model FORT" silicon probe from AppNano company was used. The resonant frequency f of the cantilever was 50-70 kHz and the stiffness factor k 1-5 N/m. Nominal tip radius was inferior to 10 nm. The AFM images were acquired in the topographic mode and analyzed with the Gwyddion software. The local measurements were performed on three different zones on each sample.

## Results and discussion

### Electrochemical conditions for Cl-induced alteration of the passive film

Several parameters such as electrolyte, pH, chloride concentration, temperature, time, polarization potential, material composition and microstructure, surface composition and microstructure, and surface roughness can influence the initiation of localized corrosion and pit growth on stainless steel [54-57]. In this work, only the polarization potential and the chloride concentration were varied in order to find the appropriate conditions to alter the passive state, however without initiating localized corrosion at least at the micrometer scale. All other parameters were kept unchanged by careful reproduction of the experimental conditions.

Figure 1a shows the polarization curve recorded in the Cl-free electrolyte in the range -0.8 V < U < 1.5 V/SCE. In the anodic region starting at -0.34 V/SCE, a narrow "active" domain is observed before passivation at -0.30 V/SCE. At the active-passive transition, the current density is quite low ($1.3 \times 10^{-5}$ A $cm^{-2}$) owing to the presence of the native oxide initially protecting the surface and already impeding dissolution before anodic passivation. The passive domain extends from -0.30 V/SCE to about 0.80 V/SCE, beyond which transpassivity is observed before the onset of oxygen evolution. In the passive domain, the lowest current density, of about $5 \times 10^{-6}$ A $cm^{-2}$, is reached in the 0.3-0.5 V/SCE range. Since corresponding to the most stable surface state reached in these non-steady state conditions, the lower limit at 0.3 V/SCE of this potential range was selected as passivation potential value for producing the steady state passive films to be studied.

Figure 1b shows the passivation current transients obtained at $U_{Pass}$ = 0.3 V/SCE and NaCl concentrations of 0 M, 0.05 M and 0.1 M. Due to instabilities of the measurement during the first second, all curves are started at 1s. The log-log plot shows that in all cases the current density continuously decreases, by about 3 orders of magnitude in the time frame of the experiments. This means that the passive films become increasingly protective with time however without reaching stable steady state after 1 hour. Previous studies with ultra-low current measurement setups and micrometer size electrodes have shown that a stable steady state is beyond reasonable time reach on austenitic stainless steels [58,59]. Current spikes, both positive and negative with respect to the passive current, are observed independently of the $Cl^-$ concentration. Some appeared during the current range changes at $10^{-4}$, $10^{-5}$, and $10^{-6}$ A $cm^{-2}$. The others correspond to instabilities generated by the electronic noise of the setup and captured by the measurement sampling (50 points per second) in the lowest current ranges. In 0.1 M NaCl, the passive current is systematically higher than in 0 M NaCl, showing that a less stable film is formed at least in the later stages of passivation ($1.6 \times 10^{-7}$ vs $0.9 \times 10^{-7}$ A $cm^{-2}$ after 3600 s) and thus a poisoning effect of chlorides on passivation.

In 0.05 M NaCl, the passive current is also higher in the later stages of passivation ($1.2 \times 10^{-7}$ vs $0.9 \times 10^{-7}$ A cm$^{-2}$ after 3600 s of passivation), showing a milder poisoning effect than in 0.1 M NaCl.

Figure 1c shows a blowup in semi-log scale of the current transients in the 1700-1800 s time range. In 0.1 M NaCl, one observes, in addition to the spike already seen in Figure 1b, a transient typical for metastable pitting. The slow increase of current represents the anodic dissolution during the initial growth of the pit, and the sudden shut down the repassivation of the pit [58-60]. No such transient are observed in 0.05 M NaCl and in 0 M NaCl. The passive currents are increasingly higher with increased chloride concentration and appear stable in this semi-log scale although they are not (Figure 1b). Figure 1d presents another blowup of the current transients corresponding to the 3500-3600 s time range. No transients associated to metastable pitting are observed in 0.05 M NaCl and in 0 M NaCl. The positive current spikes observed in 0.1 M NaCl possibly correspond to metastable events occurring too fast to be resolved by the measurement sampling. The differences in passive currents in this later stage of passivation with increasing chloride concentration confirm the poisoning of the passive state.

From the current transient of metastable pitting observed in Figure 1c, one can estimate the dimension of the associated metastable pitting event. To do so we consider that only one pit has formed in a single event and that it has a hemispherical shape. The principal reaction considered is the oxidative dissolution of the metal M(0) to M(II) with the transfer of 2 electrons. Since Fe is the base element and less noble metal in the metallic substrate, we consider that it is the only dissolved element during the metastable pitting process. According to the Pourbaix diagram, $Fe^{2+}$ is the most stable ion in the solution in our electrochemical polarization conditions of 0.3V/SCE at pH near 1.2.

The repassivation charge associated with the formation of Fe(III) and Cr(III) species is neglected since metastable pitting mostly correspond to transient dissolution. The diameter d of the pit is expressed as:

$$d = 2*\sqrt[3]{\frac{2*3*Q*M}{4*2*e*N_A*\rho*\pi}} \qquad \text{eq.1}$$

where Q is the charge of the current transient (calculated by subtracting the passive current taken as background), e the elementary charge ($1.6 \times 10^{-19}$ C), $N_A$ the Avogadro number ($6.02 \times 10^{23}$ $mol^{-1}$), M the molar mass of the metal (56 g $mol^{-1}$ for Fe) and ρ its density Fe (7.8 g $cm^{-3}$ for Fe).

From the charge of 6.1 μC of the current transient measured in 0.1 M NaCl (Figure 1c), one obtains a diameter of 9.5 μm. The presence of pits of micrometric dimensions was confirmed by optical microscopy. This means that not only metastable pitting has occurred, which we want to avoid, but also that the generated pit, or pits if several have formed simultaneously, has reached micrometer dimensions before repassivating, which is far beyond the early stage of passivity alteration preceding passivity breakdown that we want to study. In other words, the concentration of 0.1 M NaCl is too high for our study. In 0.05 M NaCl, no metastable pitting transient are detected and this concentration is thus more adapted for our investigation.

Eq. 1 can also be used to calculate the charge Q associated with the formation a single hemispherical pit of diameter d. For a pit of sub-micrometer diameter of 500 nm (considered as the lowest value for a sub-micrometric pit), the estimated charge is 0.88 nC. Assuming a metastable pitting time of 1 s, the current amplitude of the associated transient would be 0.88 nA, which is below the background currents of 80 and 60 nA measured after 3600 s in the 0.1 and 0.05 M NaCl electrolytes. The dimensions of the working electrode could be reduced to micrometer size in order to reduce the background current and detect sub-micrometer and even nanometer metastable pitting events, like achieved in previous studies [58,59]. However, this was not considered here for the sake of surface analysis that requires a measuring area of at least 1 $mm^2$. Hence, even though the 0.05 M NaCl electrolyte was selected for mild alteration of the passive state in the presence of chloride, it cannot be excluded that nanometer scale and even sub-micrometer scale metastable pits formed during the applied passivation treatment.

## Passivation transients and surface topography

Figure 2 shows the log-log plots of the passivation transients obtained in the 0.05 M $H_2SO_4$ + 0.05 M NaCl electrolyte starting from the native oxide-covered surface state and the pre-passivated surface state on the samples subsequently used for surface analysis. The transient obtained in the Cl-free 0.05 M $H_2SO_4$ electrolyte starting from the native oxide-covered surface state, and corresponding to the pre-passivation treatment, is also shown for comparison. No current spikes are observed in these experiments performed in more stable measuring conditions. Like in Figure 1, the plots comparing the samples passivated and the native oxide-covered surface state with or without chlorides in the electrolyte show a continuous decrease in passive current, confirming that an increasingly protecting passive film is formed and that the stable steady state is not reached after 1 hour. The slightly higher current measured in the 0.05 M NaCl electrolyte confirms the alterations of the passive state at the later stage of passivation after 3600 s ($1.2 \times 10^{-7}$ vs $1.0 \times 10^{-7}$ A $cm^{-2}$). A slightly higher current is also observed in the early passivation stages after 1 s ($1.8 \times 10^{-4}$ vs $1.4 \times 10^{-4}$ A $cm^{-2}$), a difference more marked than in Figure 1 possibly due to slight differences in the native oxide-covered surface state. The poisoning effect of chlorides is observed also during initial passivation in this experiment and it is continuous.

Based on the relation between volume of consumed material and associated charge, it is possible to express the equivalent thickness h dissolved during the passivation process. Still considering that the metal M(0) oxidizes to M(II) with the transfer of 2 electrons, the expressions is:

$$h = \frac{Q*M}{2*e*N_A*\rho\ *S} \qquad \text{eq.2}$$

where S is the electrode area (0.5 $cm^2$ in our case).

For the transient measured in the Cl-free 0.05 M $H_2SO_4$ electrolyte starting from the native oxide-covered surface state (black curve), the charge obtained by integrating the current curve from 1 to 3600 s is 1.9 mC corresponding to an equivalent thickness of 1.4 nm. From the *fcc* structure of stainless steel of parameter 0.359 nm, this corresponds to 7.8 atomic (100) planes of stainless steel being consumed by the transition from native-oxide surface state to passivated surface state. For the transient obtained in the Cl-containing 0.05 M $H_2SO_4$ electrolyte also starting from the native oxide-covered surface state (red curve), the charge is 2.3 mC corresponding to an equivalent thickness of 1.7 nm. This slightly higher (about 20%) value shows that the passivation process consumes more material in the presence of chlorides in the electrolyte, and thus confirms the poisoning effect of chlorides on passivation.

On the pre-passivated sample, the decrease in passive current is also continuous and a stable steady state is not reached after 3600 s. However, the current has much lower amplitude, starting from $3.5 \times 10^{-6}$ A $cm^{-2}$ after 1 s, nearly two orders of magnitude lower, and ending at $7.0 \times 10^{-8}$ A $cm^{-2}$ after 3600 s, also lower that in the absence of pre-passivation. This shows that the pre-passivation treatment increases the stability of the surface state, including in the presence of chlorides in the electrolyte, and suggests that it reduces the Cl-induced alterations of the passive state. The charge measured during this treatment is 0.4 mC corresponding to an equivalent thickness of 0.3 nm. This value, markedly lower than the value of 1.4 nm, confirms the beneficial effect of pre-passivation on the surface residual reactivity in the presence of chlorides.

Figure 3 presents the AFM topographic data for the initial native oxide-covered surface and the surface passivated in the Cl-containing electrolyte. In both cases, the grooves produced by the mechanical polishing treatment are clearly seen. Protruding micro particles are also present, likely originating from dust deposited from the ambient atmosphere. The two histograms present the distribution of the pixels along the surface normal (Z position). Each distribution is referred to a zero value corresponding to the median topographic level of the image. For the native oxide-covered surface (Figure 3a), the Z axis shows the deepest value of -15.9 nm, which is confirmed in the histogram (Figure 3c). The number of pixels at a depth in the range from -12 to -16 nm does not exceed 100 counts. They correspond to the deepest points located in the polishing grooves. For the passivated surface (Figure 3b), the deepest point with respect to the median topographic level is lower, -24 nm. From the histogram (Figure 3d), the number of pixels in the range from -12 to -16 nm now exceeds 100 counts and there are more than 60 pixels in the depth range from -16 to -24 nm. Identification of the location of these pixels shows that they correspond to two metastable pits of nanometer dimensions (400 nm) circled in the image and initiated at polishing grooves. Taking into account this time a disk pit shape of height h, the expression of charge Q becomes:

$$Q = \frac{\pi * (\frac{d}{2})^2 * h * 2 * e * N_A * \rho}{3 * M} \qquad \text{eq.3}$$

The charge associated to the observed metastable pits is 0.03 nC, which, considering a current transient of 0.02 s, amounts to a current of 1.5 nA and confirms that the metastable pits observed by AFM were undetectable in the current transients because masked by the background passive current.

Figure 4 presents the AFM topographic data for the surface pre-passivated in the Cl-free electrolyte and further passivated in the Cl-containing electrolyte. The polishing grooves remain clearly seen independently of the passivation treatment. For the sample pre-passivated in the Cl-free electrolyte (Figure 4a,c), the deepest point reaches -15.7 nm and the number of pixels in the range -12 to -16 nm does not exceed 100 counts, like for the native oxide-covered surface, showing no significant effect of the passivation treatment on the surface topology. The deepest pixels are also located in the polishing grooves. After passivation in the Cl-containing electrolyte (Figure 4b,d), the deepest pixel is at -10.8 nm, less deep than prior to passivation possibly due to a slight loss in resolution caused by tip blunting. The deepest pixels are in the range -10 to -12 nm (38 counts) and located in the polishing groove. No local pit of nanometer dimensions can be detected like observed in the absence of pre-passivation. These observations, repeated in three distinct local areas of the samples, confirm that the pre-passivation in the absence of chlorides increases the resistance to passivity breakdown and the initiation of localized corrosion as reported previously from macroscopic measurement [10,11]. They show that, in the conditions tested here, metastable pitting can be blocked at the micrometer scale and at least mitigated at the nanometer scale.

## Surface analysis of the Cl-induced alterations of the passive state

Surface analysis by ToF-SIMS and XPS was applied to the samples passivated in the presence of chlorides. The ToF-SIMS and XPS data for the native oxide-covered surface state and the pre-passivated surface state have been reported separately [46]. Figure 5 shows the ToF-SIMS analysis of the elemental in-depth distribution for the native oxide-covered surface state after passivation in the 0.05 M Cl-containing electrolyte. In Figure 5a, the intensities of selected secondary ions characteristic of the oxide film ($^{18}O^-$, $^{18}OH^-$, $Cl^-$, $CrO_2^-$, $FeO_2^-$, $NiO_2^-$ and $MoO_2^-$) and substrate ($Cr_2^-$, $Fe_2^-$ and $Ni_2^-$) are plotted in logarithmic scale versus sputtering time. Between the oxide film and metallic bulk substrate regions, there is a "modified alloy" region where Ni is enriched in agreement with previous studies on austenitic stainless steels [20-22,27-30,46,47]. The region was placed between 16 and 35 s of sputtering time, meaning that the oxide film region extends from 0 to 16 s and the metallic substrate region after 35 s. In the surface oxide region, the most predominant profiles of the oxidized metals are those of the $CrO_2^-$, $FeO_2^-$ and to a lesser extent $MoO_2^-$ ions while that of the $NiO_2^-$ ions has much lower intensity, like previously found for passivation in the Cl-free electrolyte [46,47].

Figure 5b compares the $CrO_2^-$ and $FeO_2^-$ ions profiles. The profiles do not peak at the same position, which is in agreement with the bilayer structure having iron and chromium oxides more concentrated in the outer and inner layers, respectively, like reported for passivation in the absence of chlorides [7,17,27,46,47]. The interface between outer and inner layers was positioned at 5 s which is the median sputtering position between the two intensity maxima. Figure 5c shows that the $MoO_2^-$ ions profile peaks concomitantly with the $FeO_2^-$ profile, meaning that Fe and Mo oxides are both concentrated in the outer layer, again like observed for passivation of 316L SS in the absence of chlorides [27,46,47]. Figure 5d shows that, starting from the extreme surface, the $CrO_2^-/FeO_2^-$ intensity ratio continuously increases in the outer and inner parts of the oxide film before reaching saturation in the inner layer, with values similar to that for the passive film formed in the Cl-free electrolyte [46,47].

Figure 5a also shows the depth distribution of anions. The $Cl^-$ ions profile is maximum at the topmost surface and continuously decreases in the surface oxide region, whereas the $^{18}O^-$ ions profile peaks at about 5 s of sputtering time where we position the interface between outer and inner layers. The $^{18}OH^-$ ions profile is also more intense in the outer layer than in the inner layer of the surface oxide. In Figure 6, the $Cl^-$ ions profile is superimposed to that recorded for the native oxide covered sample prior to passivation using a linear plot. This plot shows that chlorides are present at trace level in the native oxide film, most likely due to trace contaminants present in the products used for degreasing and cleaning after polishing and/or in ambient atmosphere. This trace contamination is accumulated in the outer part of the surface oxide. After passivation in the Cl-free electrolyte, there is no increase of the trace chloride contamination. However, after passivation in the Cl-containing electrolyte, more chlorides enter the passive film but mostly in the outer layer. This is consistent with most of the alterations occurring in the outer exchange layer of the surface oxide in direct contact with the Cl-containing electrolyte. The inner barrier layer of the surface oxide film appears barely affected by the penetration of chlorides.

Figure 7 shows the XPS core level spectra and their reconstruction for the native oxide-covered surface state after passivation in the 0.05 M Cl-containing electrolyte. The BE values, Full Widths at Half-Maximum (FWHM) values and relative intensities of the peak components obtained by curve fitting are compiled in Table. 1. The exact same fitting procedure as for the native oxide-covered surface before and after passivation in the Cl-free electrolyte was applied [46,47]. Seven peaks were used to fit the Cr $2p_{3/2}$ experimental curve (Figure 7a) as proposed previously [61]. The first one (Cr1) is associated to metallic $Cr^0$ in the substrate. The next five peaks (Cr2-Cr6) form a well-defined series with fixed BE intervals, FWHMs and relative intensities and correspond to $Cr^{III}$ oxide in the inner part of the oxide film. The additional peak (Cr7) is associated to $Cr^{III}$ hydroxide in the outer part of the oxide film. No $Cr^{VI}$ expected at a BE of ~579.5 eV [61] was needed for curve fitting.

Seven peaks were also used for fitting the Fe $2p_{3/2}$ spectrum [61,62] (Figure 7b). The first peak (Fe1) corresponds to metallic $Fe^0$ in the substrate, the next five peaks (Fe2-Fe6), also forming a well-defined series, correspond to $Fe^{III}$ oxide in the surface oxide layers, and the additional peak (Fe7), needed at higher BE to optimize the fit, corresponds to $Fe^{III}$ hydroxide also in the oxide film.

The Mo 3d spectrum was fitted with three 5/2-3/2 spin-orbit doublets (Figure 7c): one (Mo1/Mo1') corresponding to metallic $Mo^0$ in the substrate and the other two, (Mo2/Mo2') and (Mo3/Mo3'), to $Mo^{IV}$ and $Mo^{VI}$ in the surface oxide film, respectively [21,22,27,31,33,36,46,47,61]. The presence of an S 2s peak (S1 component) can be observed in Figure 7c (Table 1). It originates from the sulfate counter ions of the solution. The S 2p components measured at 168.8 and 170.0 eV (Figure 7e, Table 1) confirm the presence of $SO_4^{2-}$ species at the surface of the passive film.

The Ni 2p3/2 spectrum was fitted using a single peak (Table 1) corresponding to metallic $Ni^0$ in the substrate [27,46,47]. No Ni(II) oxide species were observed, like for the native oxide film and after passivation in the Cl-free electrolyte [46,47], evidencing that the Ni oxide species measured by ToF-SIMS were at trace level below the detection limit of XPS (~0.5 at%).

Figure 7d shows the O 1s spectrum fitted with three components, O1, O2 and O3, assigned to the oxide ($O^{2-}$), hydroxide ($OH^-$) and water ($H_2O$) ligands in the surface oxide films, respectively [27,46,47,61]. The $OH^-/O^{2-}$ intensity ratio is 1.5, higher than for the native oxide (0.9) but also higher than after passivation in the Cl-free electrolyte (1.2) [46], meaning that the presence of chlorides in the outer layer of the passive film is associated with an increased hydroxylation as suggested by the ToF-SIMS depth profile.

Figure 7f shows the Cl 2p core level region spectrum. No peaks are observed, meaning that the quantity of chlorides measured by ToF-SIMS was below the detection limit of XPS (~0.5 at%).

The bi-layer model previously used to calculate the thickness and composition of the outer and inner layers of the oxide films as well as the composition of the modified alloy underneath the oxide films [46] was applied to process the XPS intensity data. This model assumes a mixed iron-chromium hydroxide outer layer and a mixed iron-chromium oxide inner layer. Molybdenum oxide is included in the outer layer and neglected in the inner layer. Assignment of the intensities of the different components was as follows: Cr2-Cr6 and Fe2-Fe4 components to the oxide film inner layer, Cr7, Fe5-Fe7 and Mo2-Mo3 components to the oxide film outer layer and Cr1, Fe1, Mo1 and Ni1 components to the modified alloy region underneath the oxide film. The results are presented in Table 2, together with a reminder of the results for the native oxide-covered surface and the surface passivated in the Cl-free solution corresponding to the pre-passivation treatment. The overall compositions of the oxide films were obtained by weighting the cation concentration value of each element by the fractional thickness of the inner and outer layers.

Comparing the present data for passivation in the Cl-containing electrolyte with the native oxide-covered surface [46], one notices no significant difference in thickness of the inner layer taking into account an uncertainty of ±0.1 nm, like after passivation in the Cl-free electrolyte. The outer layer is possibly slightly thicker owing to the incorporation of chlorides and increased hydroxylation. The further enrichment of the inner layer in Cr(III) obtained by passivation in the Cl-free solution appears to be blocked after passivation in the Cl-containing electrolyte. The further enrichment in Cr(III) in the outer layer is also attenuated in the presence of chlorides. This is also observed for the enrichment in Mo(IV,VI) of the outer layer. No marked variation of the composition of the modified alloy underneath the surface oxide are observed. Thus it appears that the incorporation of chlorides in trace amounts (< 0.5 at%) and the increased hydroxylation of the outer layer of the surface oxide is detrimental to the passive state since mitigating the increase of the beneficial Cr and Mo enrichments brought by passivation, including in the inner layer where chlorides do not enter.

## Surface analysis of the effect of pre-passivation

Similarly to Figure 5, Figure 8 presents the ToF-SIMS data for the sample pre-passivated in Cl-free conditions and further treated in the Cl-containing electrolyte. The "modified alloy" region was positioned between 18 s to 37 s (Figure 8a). The bilayer structure of the oxide film is still observed with the interface between outer and inner layers placed at 6 s (Figure 8b). Cr oxide remains concentrated in the inner layer and Fe oxide in the outer layer with Mo oxide (Figure 8c). The variation of the intensity ratio of $CrO_2^-$ to $FeO_2^-$ ions, shown in Figure 8d, shows also a similar trend, confirming the higher chromium enrichment in the inner layer. A higher value is reached in the inner layer, suggesting increased enrichment in Cr oxide.

Figure 8a also shows that the depth profiles for the $Cl^-$, $^{18}O^-$ and $^{18}OH^-$ ions are similar to those for the non-pretreated sample (Figure 5a), however with different intensities. The $Cl^-$ ion profile has much lower intensity. It is at a level similar to those measured in the samples not exposed to chlorides as shown in Figure 6, indicating that the pre-passivation treatment would block the entry of the chlorides in the passive film. The $^{18}OH^-$ ions profile has also a lower intensity owing to the pre-passivation treatment in the absence of chlorides [46,47].

Similarly to Figure 7, Figure 9 presents the XPS core level spectra recorded for the sample pre-passivated in Cl-free conditions and further treated in the Cl-containing electrolyte. The spectra could be reconstructed using the same components with no significant differences of the BE and FWHM values within the accuracy of ±0.1 eV (Table 1). Only slight variations of the relative intensities were observed. In the O 1s core level region, the $OH^-/O^{2-}$ intensity ratio is 1.2, the same as for the film after pre-treatment in the Cl-free electrolyte (1.2) [46], but lower than on the sample post-treated in the Cl-containing electrolyte (1.5) thereby confirming the poisoning effect of chlorides on the dehydroxylation of the passive film. In the Cl 2p core level region (Figure 9f), no signal is detected meaning that the residual trace quantity of chlorides measured by ToF-SIMS and already observed before treatment in the Cl-containing electrolyte was below the detection limit of XPS (~0.5 at%).

The results for the calculated thickness and composition are compiled in Table 2. Compared with the data for the surface after the pre-passivation treatment, one observes no significant variations of thickness of the outer and inner layers. The inner layer is slightly more enriched in Cr(III) whereas the outer layer is less enriched in Cr(III) but more in Mo(IV,VI), which are the variations also observed to be induced by aging under anodic polarization in the absence of chlorides [46]. Thus, it is suggested that, thanks to the blocking effect of the pre-passivation treatment on the entry of chlorides in the outer layer of the passive film, the aging-induced variations of the composition could take place despite the presence of chlorides in the environment.

These results show that pre-passivation by anodic polarization in the Cl-free electrolyte increases the stability of the passive state obtained in the Cl-containing electrolyte and blocks the entry of the chlorides in the passive film, including in the outer exchange layer. Comparison of the data in Table 2 for the native oxide and pre-passivated films shows that pre-passivation in chloride-free environment does not significantly change the thickness of the passive film but that the composition of the passive film is changed, including in the outer layer further enriched in Cr and Mo by the pre-passivation process. It is suggested that this change of composition, possibly modifying the n-type electronic properties [63,64] and thus the driving force for chloride entry under anodic polarization, is at the origin of the blocking effect of pre-passivation on the chloride entry. Curing structural effects on the local sites for entry of chloride ions are suggested.

## Conclusion

Surface analysis by ToF-SIMS, XPS and AFM were combined to electrochemical measurements in order to study the alterations of the passive state of 316L austenitic stainless steel brought by the presence of chlorides in sulfuric acid passivating electrolyte and the effect of pre-passivation in the absence of chlorides. The passivation conditions (potential, time and Cl concentration) were selected so as to enable Cl-induced alterations of the passive film, however without generating stable pit growth and limiting metastable pitting to the nanometer scale as confirmed by current transients and AFM measurements.

Passivation by anodic polarization in the Cl-containing electrolyte causes $Cl^-$ ions to enter the bilayer structure of the surface oxide already formed in the native oxide-covered surface state, mostly in the hydroxide outer layer where Fe(III) and Mo(IV,VI) species are concentrated but barely in the oxide inner layer enriched in Cr(III). Their concentration is below the detection limit of XPS (< 0.5 at%). The passivation mechanisms leading to dehydroxylation of the oxide film and to further Cr(III) enrichment in the outer exchange layer and inner barrier layer and to Mo(IV,VI) enrichment in the outer layer are negatively impacted, resulting in a less protective passive state poisoned by the interactions with chlorides. These alterations suggest that the preferential dissolution of iron oxide, leading to Cr and Mo enrichment, is slow down by the interactions with the $Cl^-$ ions at the surface of the passive film.

Pre-passivation by anodic polarization in the Cl-free electrolyte increases the stability of the passive state obtained in the Cl-containing electrolyte. It blocks the entry of the chlorides in the passive film, including in the outer exchange layer, which enables the aging-induced variations of the composition to take place despite the presence of chlorides in the environment.

## Acknowledgements

This project has received funding from the European Research Council (ERC) under the European Union's Horizon 2020 research and innovation program (ERC Advanced Grant no. 741123). Région Île-de-France is acknowledged for partial funding of the ToF-SIMS equipment.

## Figures captions

*Figure 1. Electrochemical polarization of the native oxide-covered 316L SS sample in 0.05 M $H_2SO_4$: a) Potentiodynamic polarization curve in the Cl-free electrolyte (scan rate 1 mV/s), b) Potentiostatic polarisation curves at 0.3 V/SCE at increasing chloride concentration, c) Zoom of (b) in the 1700-1800 s time range, d) Zoom of (b) in the 3500-3600 s time range.*

*Figure 2. Potentiostatic polarisation curves at 0.3 V/SCE for 316L SS in 0.05 M $H_2SO_4$ for the native oxide-covered sample in the Cl-free electrolyte (Nat Ox; 0 M $Cl^-$) and in the 0.05 M NaCl electrolyte (Nat Ox; 0.05 M $Cl^-$), and for the pre-passivated sample in the 0.05 M NaCl electrolyte (Pre-Pass; 0.05 M $Cl^-$).*

*Figure 3. AFM analysis for the native oxide-covered sample before (a,c) and after (b,d) passivation at 0.3 V/SCE in the 0.05 M $H_2SO_4$ + 0.05 M NaCl electrolyte: (a,b) AFM images (10 × 10 $\mu m^2$), (c,d) Z position histograms.*

*Figure 4. AFM analysis for the pre-passivated sample before (a,c) and after (b,d) passivation at 0.3 V/SCE in the 0.05 M $H_2SO_4$ + 0.05 M NaCl electrolyte: (a,b) AFM images (10 × 10 $\mu m^2$), (c,d) Z position histograms.*

*Figure 5. ToF-SIMS negative ions depth profiles for 316L SS passivated at 0.3 V/SCE in 0.05 M $H_2SO_4$ + 0.05 M NaCl starting from the native oxide-covered surface state: (a) $^{18}O^-$, $^{18}OH^-$, $Cl^-$, $CrO_2^-$, $FeO_2^-$, $NiO_2^-$, $MoO_2^-$, $Cr_2^-$, $Fe_2^-$ and $Ni_2^-$ secondary ions, (b) $CrO_2^-$ and $FeO_2^-$ secondary ions, (c) $FeO_2^-$ and $MoO_2^-$ secondary ions, (d) $CrO_2^-$/ $FeO_2^-$ intensity ratio.*

*Figure 6. ToF-SIMS $Cl^-$ ions depth profiles for the native oxide-covered sample passivated at 0.3 V/SCE in the Cl-free 0.05 M $H_2SO_4$ electrolyte (Nat Ox; 0 M $Cl^-$) and in the 0.05 M NaCl + 0.05 M $H_2SO_4$ electrolyte (Nat Ox; 0.05 M $Cl^-$), and for the pre-passivated sample passivated at 0.3 V/SCE in the 0.05 M NaCl + 0.05 M $H_2SO_4$ electrolyte (Pre-Pass; 0.05 M $Cl^-$).*

*Figure 7. XPS core level spectra and their reconstruction for 316L SS passivated at 0.3 V/SCE in 0.05 M $H_2SO_4$ + 0.05 M NaCl starting from the native oxide-covered surface state: (a) Cr $2p_{3/2,}$ (b) Fe $2p_{3/2,}$ (c) Mo $3d_{5/2-3/2}$ (d) O 1s (e) S 2p and (f) Cl 2p regions.*

*Figure 8. ToF-SIMS negative ions depth profiles for 316L SS passivated at 0.3 V/SCE in 0.05 M* $H_2SO_4$ *+ 0.05 M NaCl starting from the pre-passivated surface state: (a)* $^{18}O^-$, $^{18}OH^-$, $Cl^-$, $CrO_2^-$, $FeO_2^-$, $NiO_2^-$, $MoO_2^-$, $Cr_2^-$, $Fe_2^-$ *and* $Ni_2^-$ *secondary ions, (b)* $CrO_2^-$ *and* $FeO_2^-$ *secondary ions, (c)* $FeO_2^-$ *and* $MoO_2^-$ *secondary ions, (d)* $CrO_2^-/FeO_2^-$ *intensity ratio.*

*Figure 9. XPS core level spectra and their reconstruction for 316L SS passivated at 0.3 V/SCE in 0.05 M* $H_2SO_4$ *+ 0.05 M NaCl starting from the pre-passivated surface state: (a) Cr* $2p_{3/2}$*, (b) Fe* $2p_{3/2}$*, (c) Mo* $3d_{5/2-3/2}$ *and (d) O 1s (e) S 2p and (f) Cl 2p regions.*

## Tables captions

*Table 1. BE, FWHM and relative intensity values of the peak components measured by XPS on 316L SS for the native oxide-covered sample passivated in the 0.05 M NaCl + 0.05M* $H_2SO_4$ *electrolyte and for the pre-passivated sample passivated in the 0.05 M NaCl + 0.05M* $H_2SO_4$ *electrolyte.*

*Table 2. Thickness and composition of the surface oxide for the 316L SS native oxide-covered sample before (from [46]) and after passivation in the 0.05 M NaCl + 0.05M* $H_2SO_4$ *electrolyte, and for the pre-passivated sample before (from [46]) and after passivation in the 0.05 M NaCl + 0.05M* $H_2SO_4$ *electrolyte.*

## Table 1

| | | | Native oxide-covered surface passivated in 0.05 M [Cl$^-$] | | | Pre-passivated surface passivated in 0.05 M [Cl$^-$] | | |
|---|---|---|---|---|---|---|---|---|
| **Core level** | Peak | Assignment | BE (±0.1 eV) | FWHM (±0.1 eV) | Intensity (%) | BE (±0.1 eV) | FWHM (±0.1 eV) | Intensity (%) |
| **Fe 2p3/2** | Fe1 | $Fe^0$ (met) | 707.0 | 0.8 | 57.9 | 707.0 | 0.8 | 63.3 |
| | Fe2 | $Fe^{III}$ (ox) | 708.8 | 1.3 | 8.3 | 708.8 | 1.2 | 8.1 |
| | Fe3 | $Fe^{III}$ (ox) | 709.8 | 1.1 | 7.5 | 709.8 | 1.1 | 7.3 |
| | Fe4 | $Fe^{III}$ (ox) | 710.7 | 1.1 | 5.8 | 710.7 | 1.1 | 5.7 |
| | Fe5 | $Fe^{III}$ (ox) | 711.7 | 1.3 | 3.3 | 711.7 | 1.2 | 3.2 |
| | Fe6 | $Fe^{III}$ (ox) | 712.8 | 2.0 | 3.3 | 712.8 | 2.0 | 3.2 |
| | Fe7 | $Fe^{III}$ (hyd) | 711.9 | 2.7 | 13.9 | 711.9 | 2.7 | 9.2 |
| **Cr 2p3/2** | Cr1 | $Cr^0$ (met) | 574.0 | 1.1 | 22.3 | 574.0 | 1.0 | 20.7 |
| | Cr2 | $Cr^{III}$ (ox) | 576.1 | 1.4 | 18.5 | 576.2 | 1.5 | 21.8 |
| | Cr3 | $Cr^{III}$ (ox) | 577.2 | 1.4 | 17.9 | 577.2 | 1.5 | 21.1 |
| | Cr4 | $Cr^{III}$ (ox) | 577.9 | 1.4 | 9.8 | 578.0 | 1.5 | 11.5 |
| | Cr5 | $Cr^{III}$ (ox) | 578.9 | 1.4 | 4.1 | 579.0 | 1.5 | 4.8 |
| | Cr6 | $Cr^{III}$ (ox) | 579.3 | 1.4 | 2.6 | 579.4 | 1.5 | 3.1 |
| | Cr7 | $Cr^{III}$ (hyd) | 577.2 | 2.5 | 24.8 | 577.2 | 2.5 | 17.1 |
| **Ni 2p3/2** | Ni1 | $Ni^0$ (met) | 852.9 | 0.9 | 100 | 852.9 | 0.9 | 100 |
| **Mo 3d5/2** | Mo1 | $Mo^0$ (met) | 227.7 | 0.5 | 27.2 | 277.7 | 0.5 | 26.7 |
| | Mo2 | $Mo^{IV}$ (ox) | 229.5 | 0.9 | 4.3 | 229.4 | 0.9 | 4.2 |
| | Mo3 | $Mo^{VI}$ (ox) | 232.5 | 2.4 | 28.8 | 232.5 | 2.5 | 29.4 |
| **Mo 3d3/2** | Mo1' | $Mo^0$ (met) | 230.8 | 0.8 | 17.9 | 230.8 | 0.8 | 17.6 |
| | Mo2' | $Mo^{IV}$ (ox) | 232.7 | 0.9 | 2.8 | 232.6 | 0.9 | 2.7 |
| | Mo3' | $Mo^{VI}$ (ox) | 235.7 | 2.4 | 19.0 | 235.6 | 2.5 | 19.4 |
| **O 1s** | O1 | $O^{2-}$ | 530.2 | 1.2 | 33.8 | 530.2 | 1.2 | 37.2 |
| | O2 | $OH^-$ | 531.7 | 1.8 | 51.1 | 531.7 | 1.7 | 43.2 |
| | O3 | $H_2O$ | 532.6 | 2.5 | 15.1 | 532.6 | 2.5 | 19.6 |
| **S 2s** | S1 | $SO_4^{2-}$ | 232.9 | 1.7 | 100 | 232.9 | 1.7 | 100 |
| **S 2p3/2** | S2 | $SO_4^{2-}$ | 168.8 | 1.1 | 66.5 | 168.7 | 1.1 | 61.6 |
| **S 2p1/2** | S3 | $SO_4^{2-}$ | 170.0 | 1.1 | 35.5 | 169.9 | 1.1 | 38.4 |

## Table 2

| | | Global film | Outer layer | Inner layer | Modified alloy |
|---|---|---|---|---|---|
| **Native oxide-covered surface [46]** | d (nm) | 2.0 | 0.8 | 1.2 | / |
| | [Fe] (at%) | 39 | 41 | 36 | 52 |
| | [Cr] (at%) | 55 | 44 | 64 | 18 |
| | [Ni] (at%) | / | / | / | 28 |
| | [Mo](at%) | 6 | 15 | / | 3 |
| | Ratio Cr/Fe | 1.4 | 1.1 | 1.8 | / |
| **Native oxide-covered surface passivated in 0.05 M $H_2SO_4$ + 0.05 M NaCl** | d (nm) | 2.0 | 0.9 | 1.1 | / |
| | [Fe] (at%) | 36 | 36 | 36 | 52 |
| | [Cr] (at%) | 57 | 48 | 64 | 18 |
| | [Ni] (at%) | / | / | / | 26 |
| | [Mo](at%) | 7 | 16 | / | 4 |
| | Ratio Cr/Fe | 1.6 | 1.3 | 1.8 | / |
| **Pre-passivated surface [46]** | d (nm) | 1.9 | 0.7 | 1.2 | / |
| | [Fe] (at%) | 26 | 26 | 26 | 51 |
| | [Cr] (at%) | 67 | 56 | 74 | 20 |
| | [Ni] (at%) | / | / | / | 26 |
| | [Mo](at%) | 7 | 18 | / | 4 |
| | Ratio Cr/Fe | 2.6 | 2.2 | 2.8 | / |
| **Pre-passivated surface passivated in 0.05 M $H_2SO_4$ + 0.05M NaCl** | d (nm) | 1.9 | 0.6 | 1.3 | / |
| | [Fe] (at%) | 26 | 34 | 22 | 51 |
| | [Cr] (at%) | 68 | 45 | 78 | 17 |
| | [Ni] (at%) | / | / | / | 29 |
| | [Mo](at%) | 6 | 21 | / | 3 |
| | Ratio Cr/Fe | 2.6 | 1.3 | 3.5 | / |

**Figure 1**

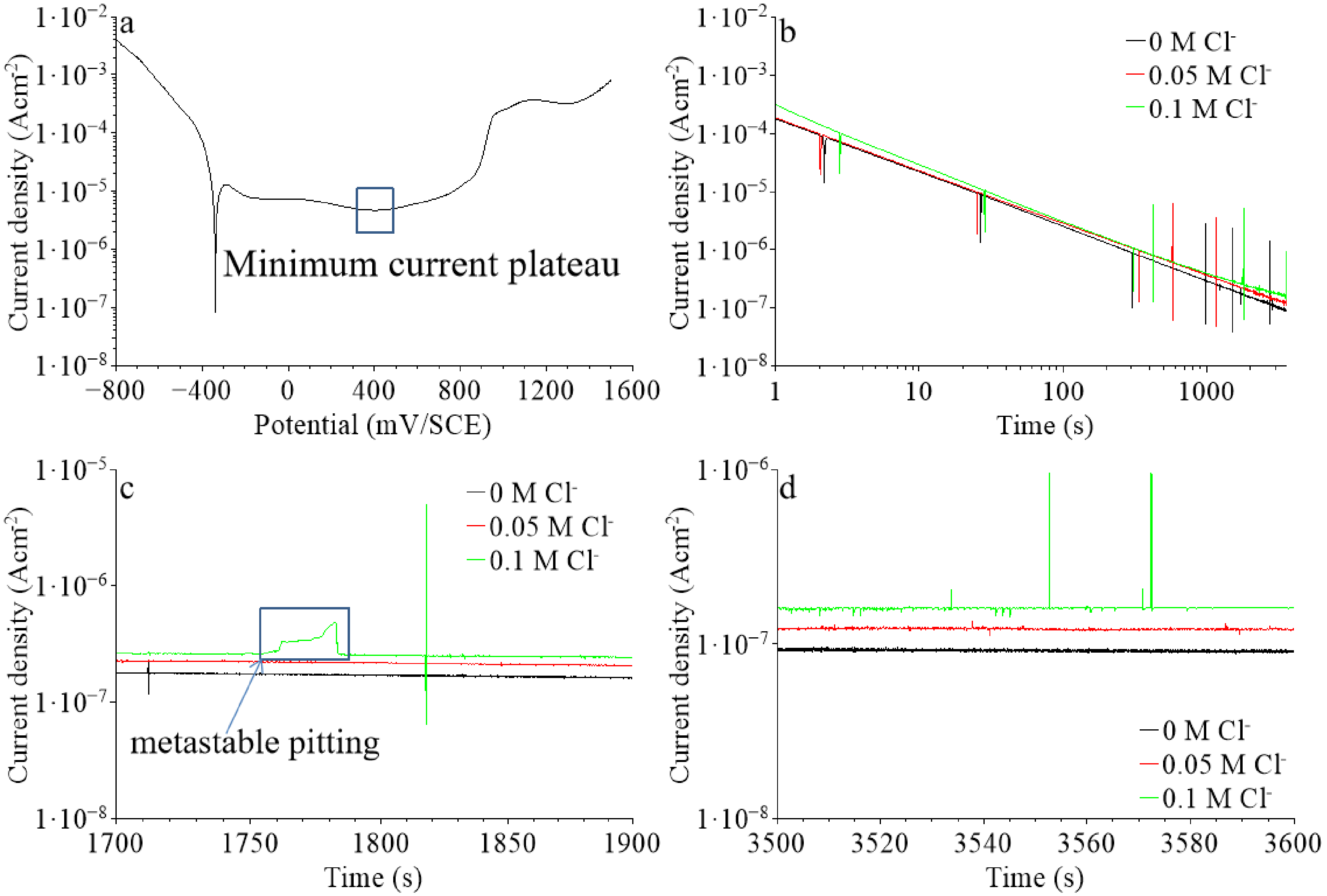

**Figure 2**

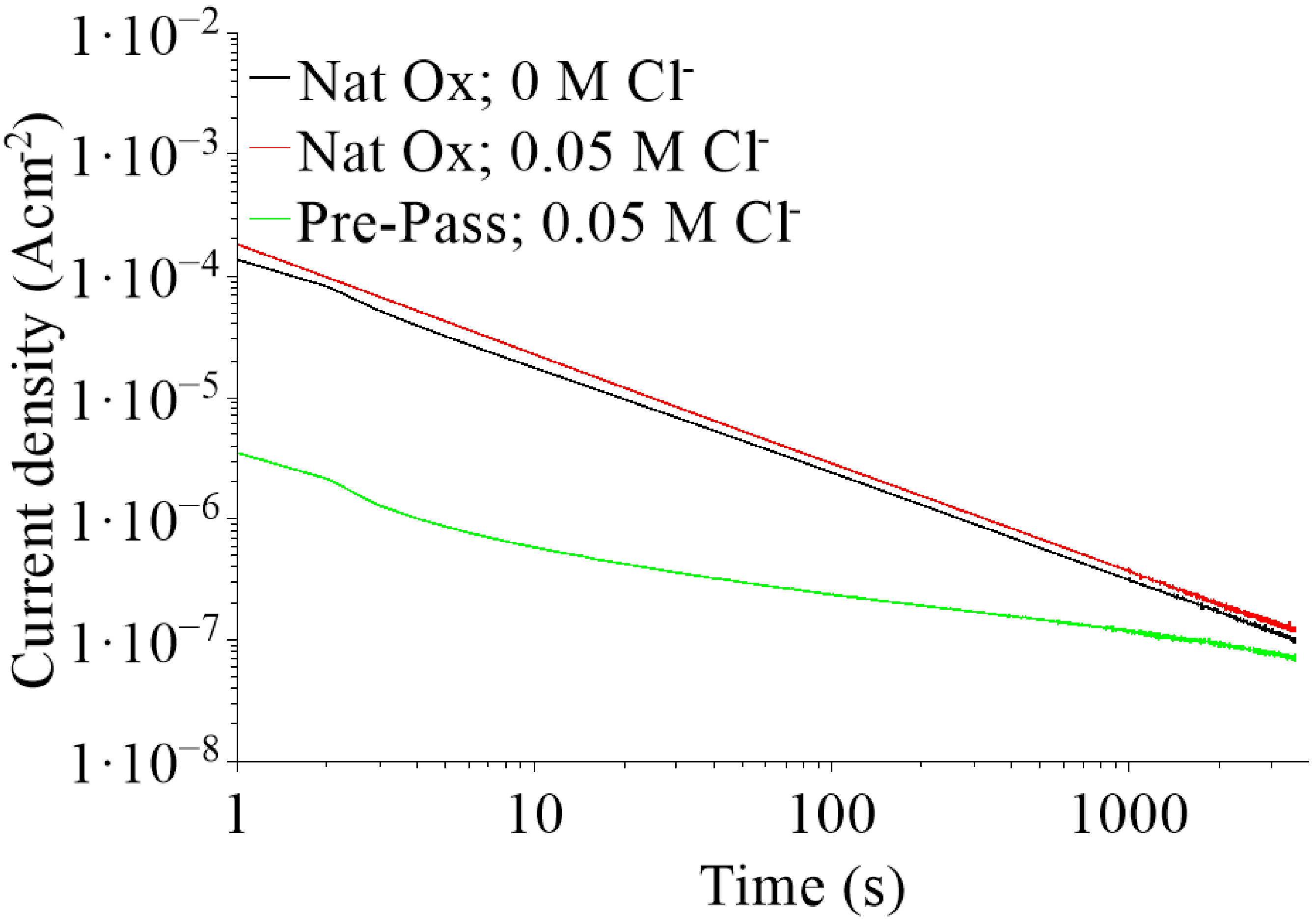

**Figure 3**

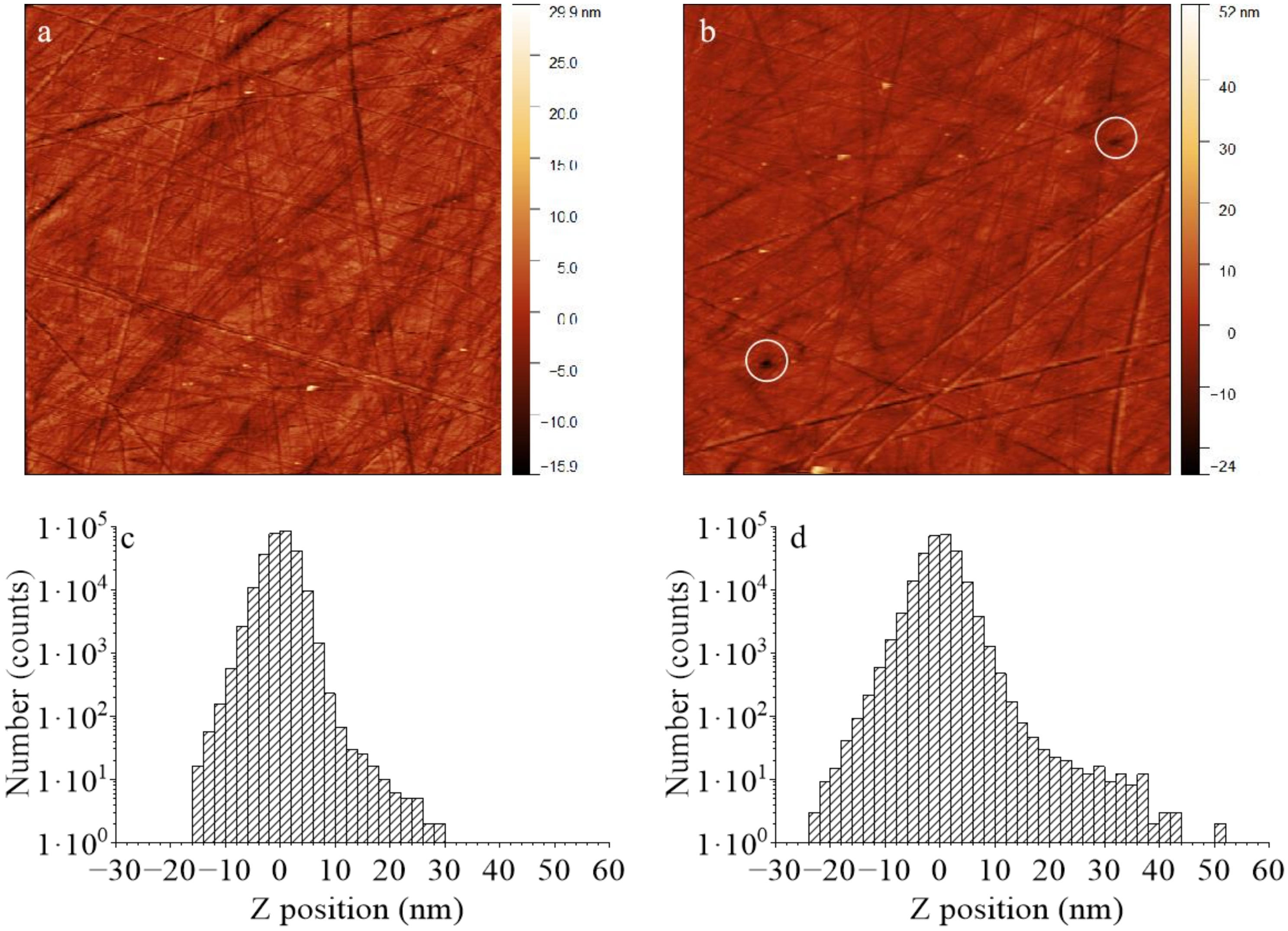

a
29.9 nm
25.0
20.0
15.0
10.0
5.0
0.0
−5.0
−10.0
−15.9
b
52 nm
40
30
20
10
0
−10
−24
c
d
Number (counts)
1·10^5
1·10^4
1·10^3
1·10^2
1·10^1
1·10^0
−30 −20 −10 0 10 20 30 40 50 60
Z position (nm)

## Figure 4

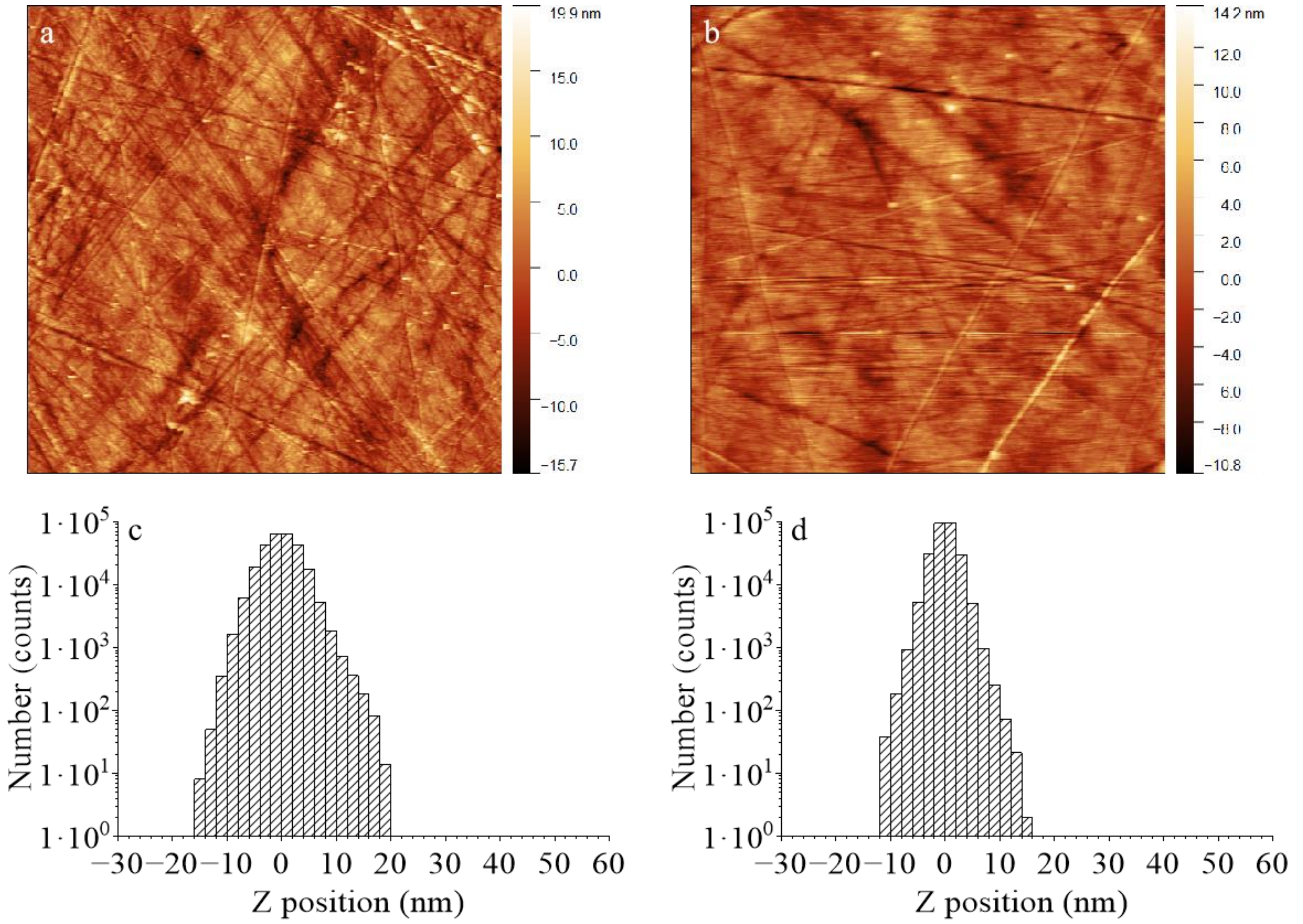
a
19.9 nm
15.0
10.0
5.0
0.0
−5.0
−10.0
−15.7
b
14.2 nm
12.0
10.0
8.0
6.0
4.0
2.0
0.0
−2.0
−4.0
6.0
−8.0
−10.8
c
Number (counts)
1·10^5
1·10^4
1·10^3
1·10^2
1·10^1
1·10^0
−30 −20 −10 0 10 20 30 40 50 60
Z position (nm)
d
Number (counts)
1·10^5
1·10^4
1·10^3
1·10^2
1·10^1
1·10^0
−30 −20 −10 0 10 20 30 40 50 60
Z position (nm)

**Figure 5**

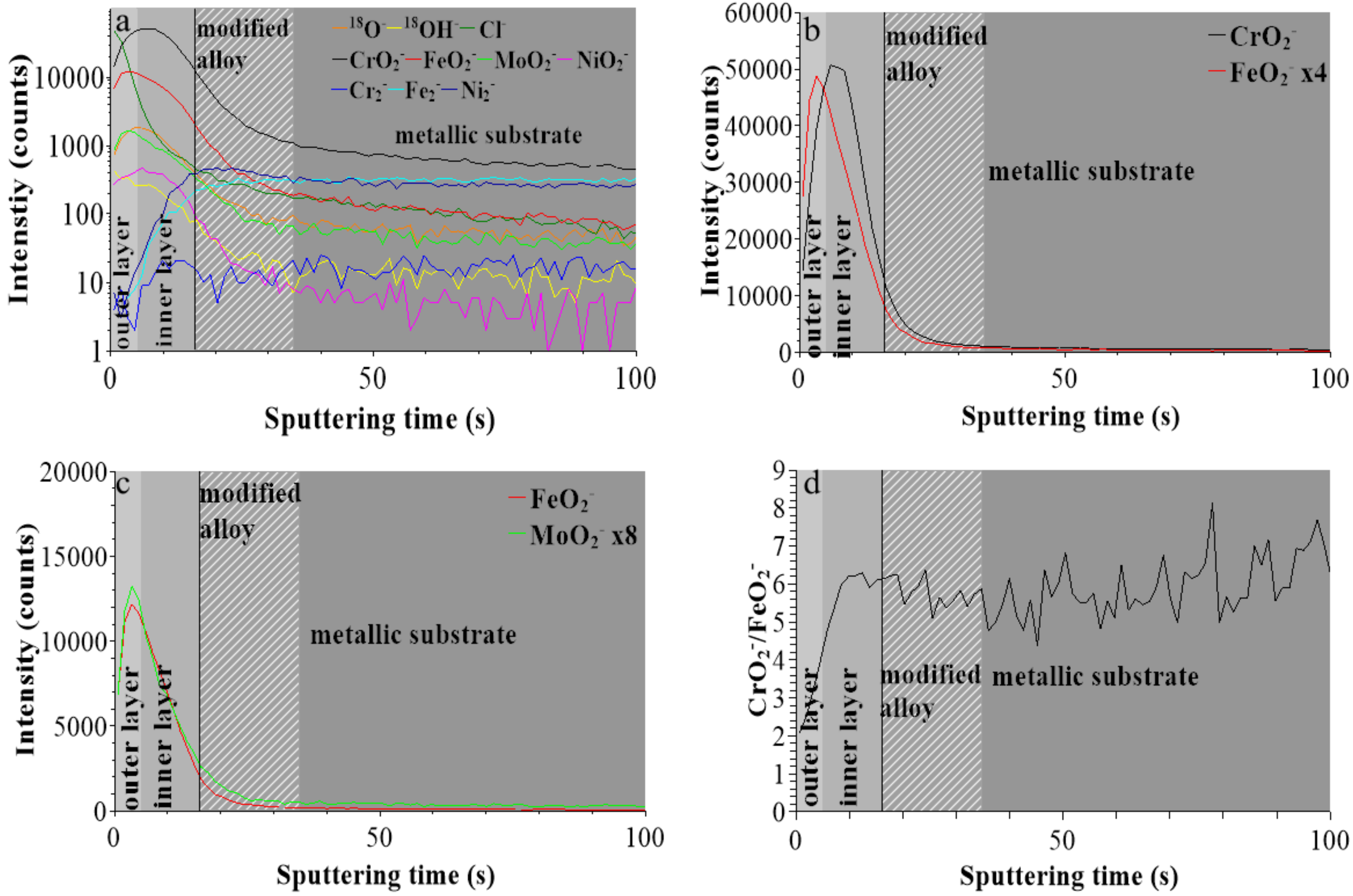

**Figure 6**

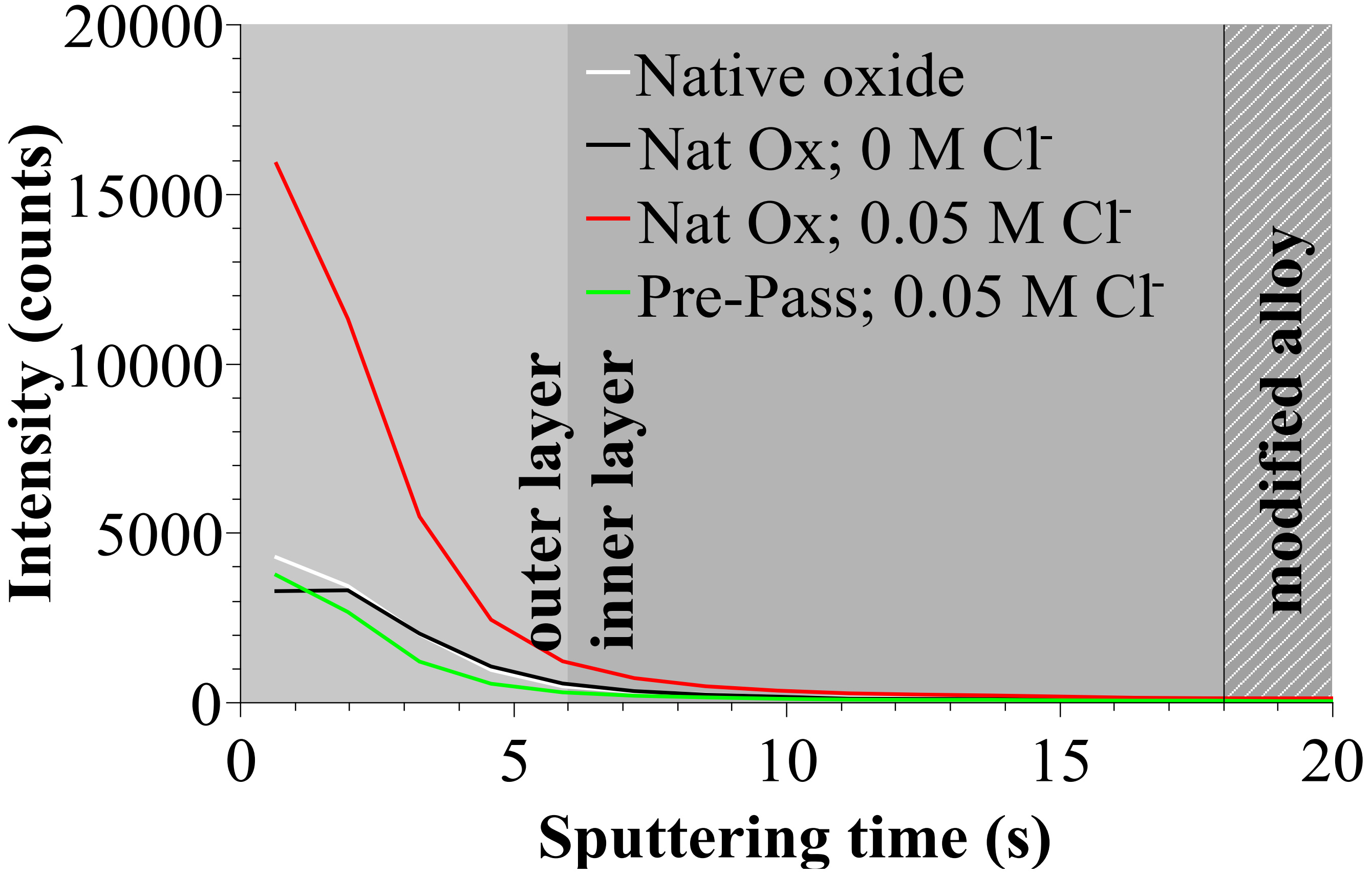

## Figure 7

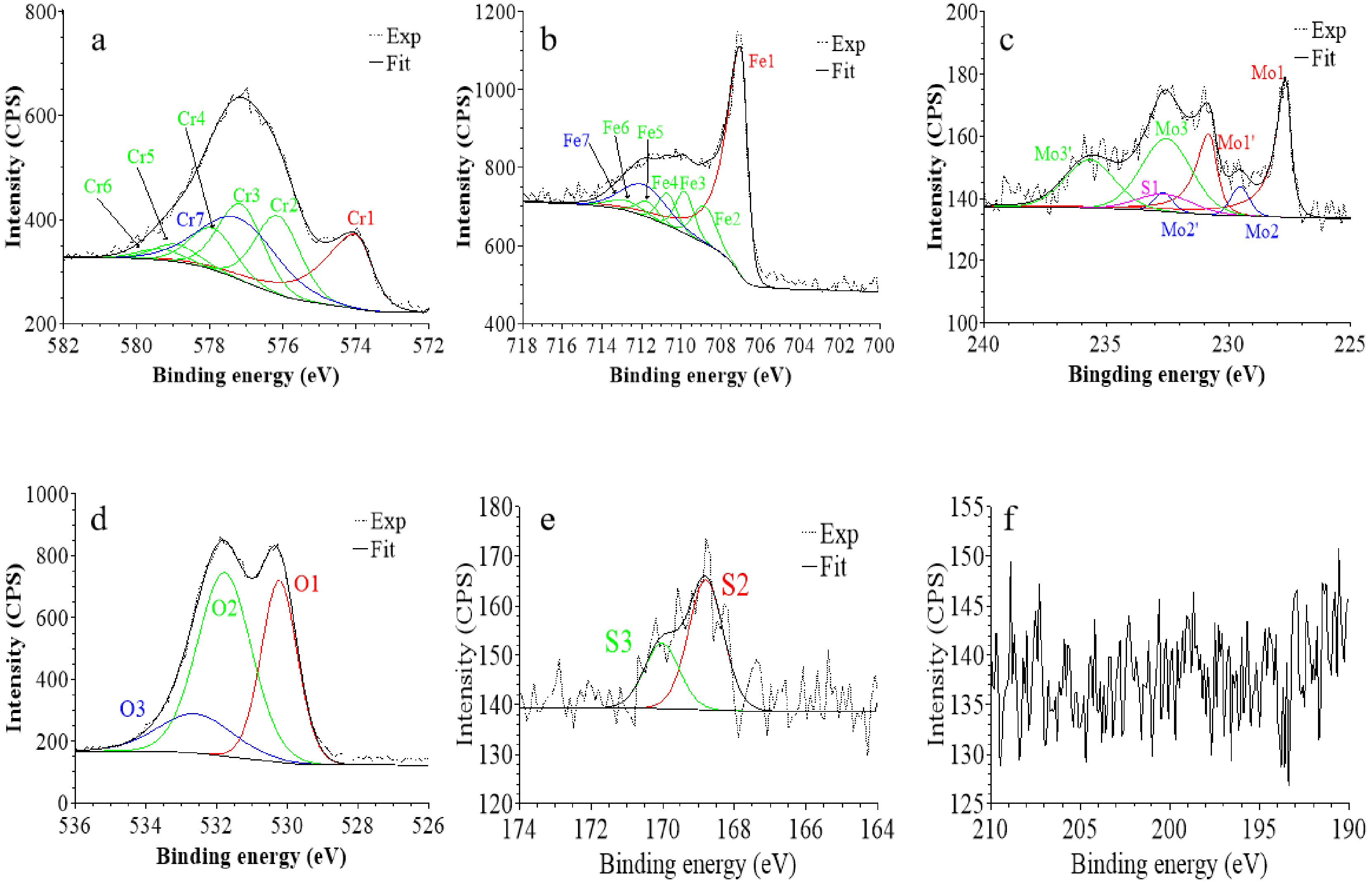

## Figure 8

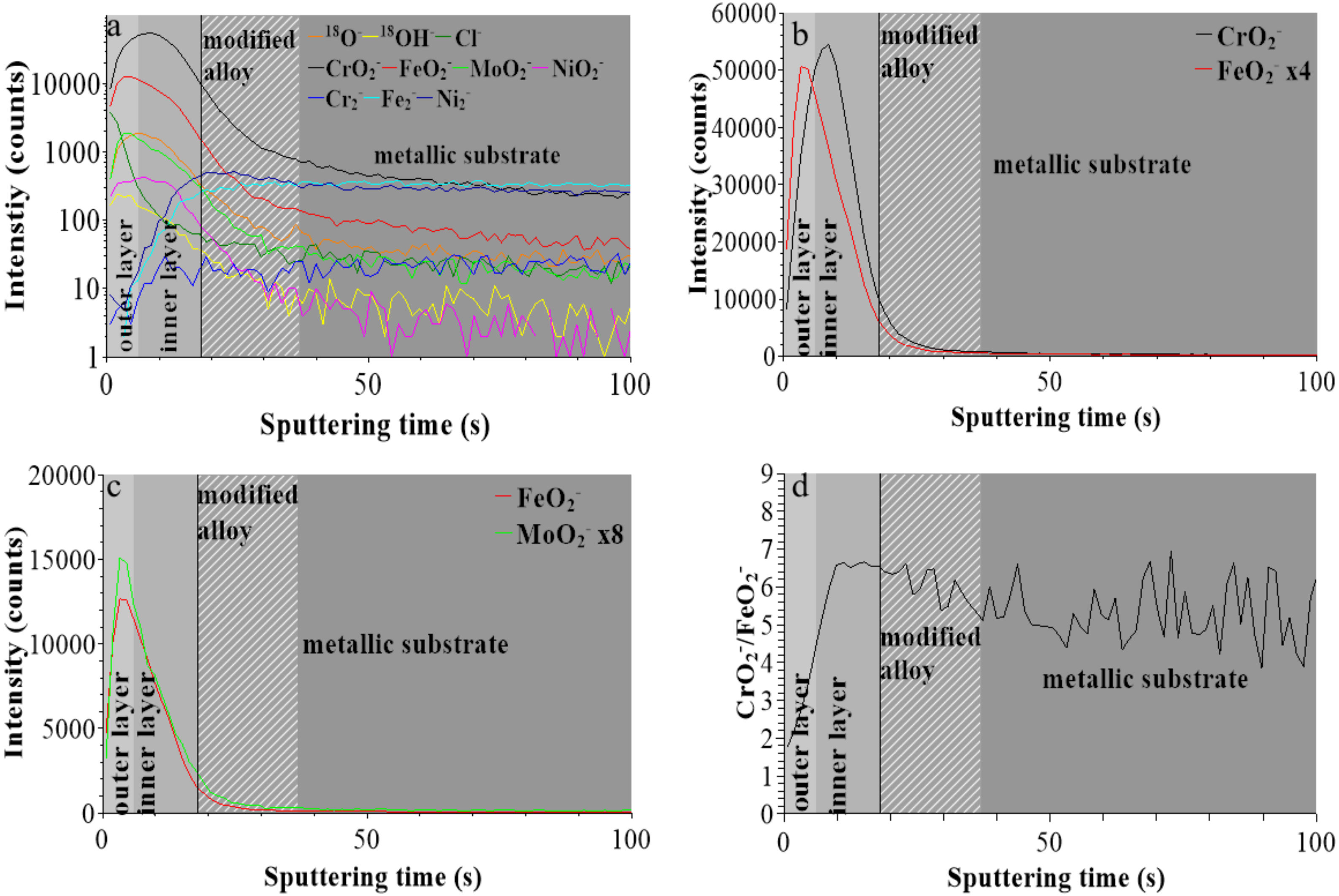

## Figure 9

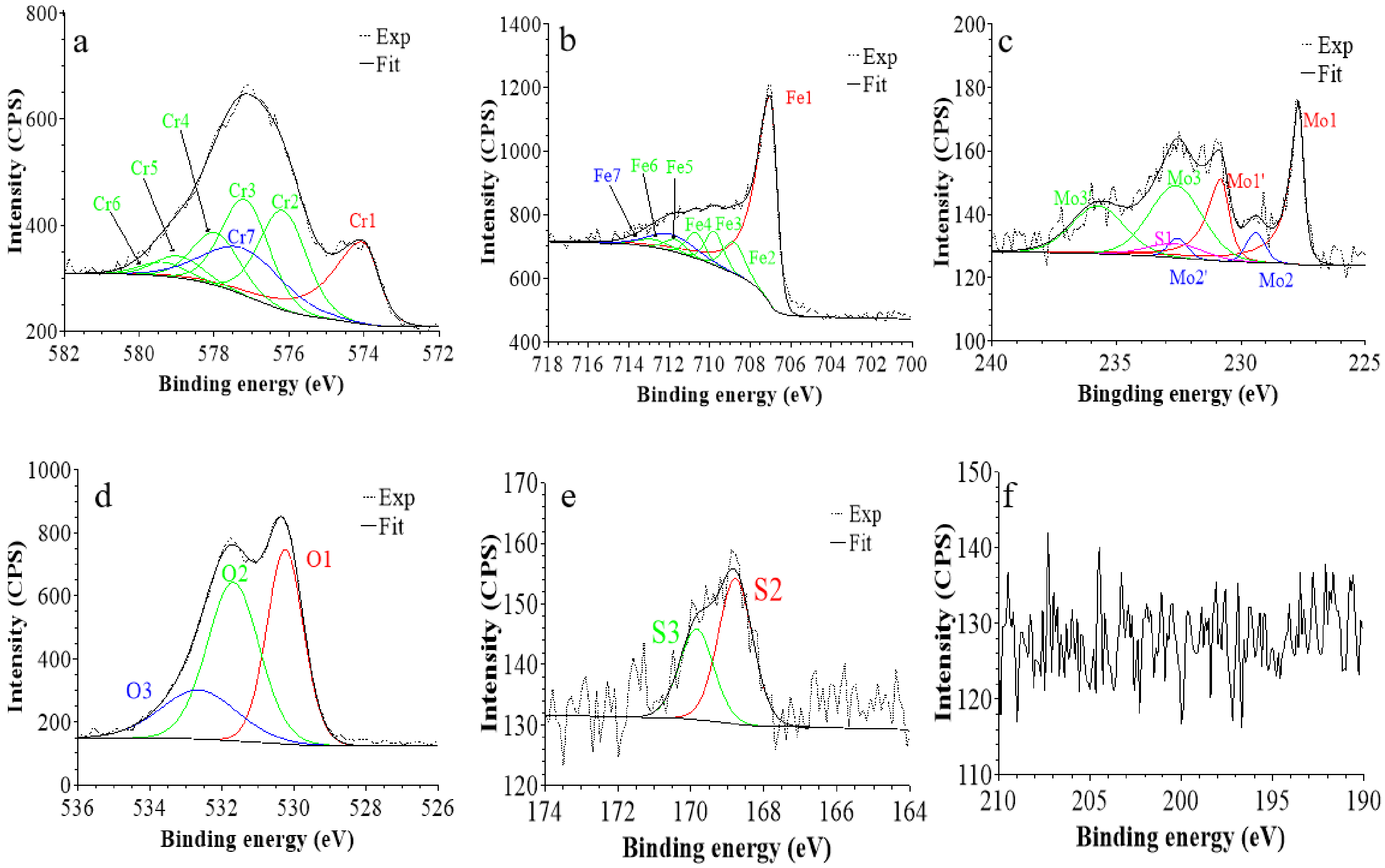